\documentclass[epj]{svjour}
\usepackage{amsmath,amssymb}
\usepackage{graphicx}
\usepackage{cite}
\def\vec#1{\mbox{\boldmath $#1$}}

\newcommand{\tr}[1]{{}^t\hspace{-0.5mm}#1}

\begin{document}
%
\title{Mechanical Responses and Stress Fluctuations of a Supercooled Liquid in a Sheared Non-Equilibrium State}
\author{Hideyuki Mizuno\inst{1,2}\thanks{E-mail:
\email{h-mizuno@cheme.kyoto-u.ac.jp}}
\and Ryoichi Yamamoto\inst{1,2}\thanks{E-mail:
\email{ryoichi@cheme.kyoto-u.ac.jp}}}
\institute{
Department of Chemical Engineering, Kyoto University, Kyoto 615-8510, Japan \and
CREST, Japan Science and Technology Agency, Kawaguchi 332-0012, Japan
}
\date{Received: date / Revised version: date}
%
\abstract{
A steady shear flow can drive supercooled liquids into a non-equilibrium state.
Using molecular dynamics simulations under steady shear flow superimposed with oscillatory shear strain for a probe, non-equilibrium mechanical responses are studied for a model supercooled liquid composed of binary soft spheres.
We found that even in the strongly sheared situation, the supercooled liquid exhibits surprisingly isotropic responses to oscillating shear strains applied in three different components of the strain tensor.
Based on this isotropic feature, we successfully constructed a simple two-mode Maxwell model that can capture the key features of the storage and loss moduli, even for highly non-equilibrium state.
Furthermore, we examined the correlation functions of the shear stress fluctuations, which also exhibit isotropic relaxation behaviors in the sheared non-equilibrium situation.
In contrast to the isotropic features, the supercooled liquid additionally demonstrates anisotropies in both its responses and its correlations to the shear stress fluctuations.
Using the constitutive equation (a two-mode Maxwell model), we demonstrated that the anisotropic responses are caused by the coupling between the oscillating strain and the driving shear flow.
Due to these anisotropic responses and fluctuations, the violation of the fluctuation-dissipation theorem (FDT) is distinct for different components.
We measured the magnitude of this violation in terms of the effective temperature.
It was demonstrated that the effective temperature is notably different between different components, which indicates that a simple scalar mapping, such as the concept of an effective temperature, oversimplifies the true nature of supercooled liquids under shear flow.
An understanding of the mechanism of isotropies and anisotropies in the responses and fluctuations will lead to a better appreciation of these violations of the FDT, as well as certain consequent modifications to the concept of an effective temperature.
}
\authorrunning{H. Mizuno and R. Yamamoto}
\titlerunning{Mechanical Responses and Stress Fluctuations of a Supercooled Liquid}
\maketitle
%
\section{Introduction} \label{sec:1}
A comprehensive theory of systems driven into non-equilibrium states is still under construction, in contrast to the well-established descriptions of equilibrium systems.
Non-equilibrium states are generally characterized by violations of the fluctuation-dissipation theorem (FDT).
The FDT relates the response functions to the associated correlation functions and holds in equilibrium states but is typically violated for non-equilibrium states.
Much work has been devoted to understanding the relationship between response functions and correlation functions in non-equilibrium situations; however, this relationship remains unclear \cite{harada_2005,speck_2006,chetrite_2009,baiesi_2009,baiesi_2010,uneyama_2011}.

It has been reported that supercooled liquids exhibit simple features even in non-equilibrium states.
A steady shear flow can drive supercooled liquids into a non-equilibrium state.
Even in strongly sheared non-equilibrium states, the structure and relaxation dynamics captured via the two-point correlation function exhibit very little anisotropy \cite{yamamoto1_1998,miyazaki_2004,bessel_2007}.
This observation is in marked contrast to observations of other complex fluids, such as polymer solutions and non-dense colloidal suspensions \cite{colloid}, in which structural changes or anisotropic dynamics are induced by a driving shear flow \cite{rheology,phasetransition}.
Furthermore, in glassy systems, including supercooled liquids, it has been suggested that the equilibrium form of the FDT holds at long times with the temperature $T$ replaced by an {effective temperature} $T_{\text{eff}}$ \cite{berthier_2000}, which indicates that $T_{\text{eff}}$ can be used to relate the response and correlation functions.
Several numerical and theoretical works have examined the validity and the role of the effective temperature in such situations \cite{berthier_2002,makse_2002,ono_2002,ohern_2004,potiguar_2006,haxton_2007,kruger_2010,zhang_2011}.

Motivated by the above reports regarding the simple non-equilibrium properties of supercooled liquids, we investigated the mechanical responses and the shear stress fluctuations of a supercooled liquid in a non-equilibrium state by means of molecular dynamics (MD) simulations.
We first drove the supercooled liquid into a non-equilibrium state by applying a steady shear flow, and we then examined the shear stress responses to oscillating shear strains in the sheared non-equilibrium state.
In this study, we considered not only the weakly sheared situation but also the strongly sheared situation.
In addition to the mechanical responses, the correlation functions of the shear stress fluctuations were also investigated in the non-equilibrium situation.
We demonstrated the violation of the FDT and measured the magnitude of this violation using the effective temperature.
Shear stress responses and fluctuations are often useful for investigating non-equilibrium statistical mechanics \cite{uneyama_2011}.
The aim of this study was to reveal the behaviors of the shear stress responses and fluctuations of the supercooled liquid in the non-equilibrium state.

Several theoretical approaches addressing the mechanical responses of glassy systems are noteworthy, including the soft glassy rheology model \cite{sollich_1998}, the shear-transformation-zone theory \cite{bouchbinder_2011_2}, and the mode-coupling theory \cite{miyazaki_2006,brader_2010,farage_2011}.
In Ref. \cite{farage_2011}, the superposition rheology of glassy materials was investigated using the mode-coupling theory.
Furthermore, in the field of complex fluid rheology, several experimental studies have examined the mechanical properties of polymer solutions under a steady shear flow \cite{yamamoto_1971,wong_1989,vermant_1998,somma_2007,li_2010}.
More recently, Ref. \cite{ovarlez_2010} performed such an experimental study for glassy materials.
We also note that constitutive equations, which detail the relationships between the stress tensor and the strain tensor, play an important role in predicting the fluid dynamics or the transport phenomena of the materials \cite{polymetric,transport}.
Several constitutive equations have been proposed to describe the mechanical properties of polymers \cite{polymetric,transport}.
To the best of our knowledge, constitutive equations for supercooled liquids (or glassy systems) have not yet been proposed for general shear strains (in tensor form).
In this study, we attempted to construct a constitutive equation for supercooled liquids that describes our simulation results.

The present paper is organized as follows.
In Sect.~\ref{sec:2}, we briefly review our MD simulation.
We also describe how to apply a steady shear flow and an oscillating shear strain.
In Sect.~\ref{sec:3} and \ref{sec:4}, the results of the mechanical responses and the stress fluctuations are presented.
In Sect.~\ref{sec:3}, we first indicate the mechanical responses obtained from the MD simulations.
In this section, we also present a constitutive equation to describe our simulation results.
In Sect.~\ref{sec:4}, we next show the results of the correlation functions of the shear stress fluctuations.
Furthermore, we demonstrate the violation of the FDT and present the effective temperature as a metric for measuring the magnitude of this violation.
Finally, in Sect.~\ref{sec:5}, we summarize our results.

\section{Simulation method} \label{sec:2}

\subsection{Simulation model} \label{sec:2-1}
In this work, we performed MD simulations in three dimensions.
Our model liquid is a mixture of two atomic species, 1 and 2, with $N_1=N_2=5,000$ particles.
The particles interact via a soft-sphere potential $\phi(r)= \epsilon (\sigma_{a b}/r)^{12}$ with $\sigma_{a b} = (\sigma_a + \sigma_b)/2$, where $r$ is the distance between two particles, $\sigma_a$ is the particle size, and $a,b \in 1,2$.
The interaction was truncated at $r=3 \sigma_{a b}$.
The mass ratio was set as $m_2/m_1=2$, and the size ratio was set as $\sigma_2/\sigma_1=1.2$ to avoid system crystallization.
Distances, times, and temperatures were measured in units of $\sigma_1$, $\tau_0 = (m_1 \sigma_1^2/\epsilon)^{1/2}$, and $\epsilon/k_B$, respectively.
The particle density was fixed at a value of $\rho = (N_1 + N_2)/V = 0.8$.
The temperature was set to be $T=0.352-0.267$.
In this study, we mainly considered the state at the temperature $T=0.306$.
Note that the freezing point of the corresponding one-component model is approximately $T=0.772$ \cite{miyagawa_1991}.
At $T=0.352-0.267$, the system is in a supercooled liquid state.
After the system was carefully equilibrated under the canonical conditions, we applied a steady shear flow and an oscillating shear strain on the system using the Lees-Edwards boundary condition \cite{nonequ}.
We integrated the SLLOD equations of motion with the Lees-Edwards boundary condition, and the temperature was maintained by the Gaussian constraint thermostat \cite{nonequ}.
The details of this simulation model can be found in previous studies \cite{yamamoto1_1998,miyazaki_2004}.

\subsection{Steady shear flow} \label{sec:2-2}
\begin{figure}
\begin{center}
\includegraphics[scale=1]{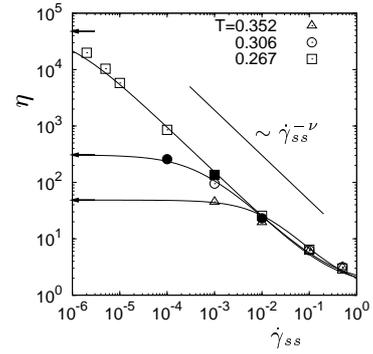}
\end{center}
\vspace*{0mm}
\caption{
The shear viscosity $\eta$ versus the shear rate $\dot{\gamma}_{ss}$ at various temperatures.
The lines are fitted to the function $\eta = \eta_{s0} /(1+\mu \dot{\gamma}_{ss}^\nu) + \eta_f$.
The arrow indicates the value $\eta_{s0} + \eta_f$, i.e., the value in the equilibrium state $\dot{\gamma}_{ss} =0$.
In this study, we primarily focused on the sheared non-equilibrium states at $T=0.306$ and $\dot{\gamma}_{ss} = 10^{-4}$ (weakly sheared state) and $T=0.306$ and $\dot{\gamma}_{ss} = 10^{-2}$ (strongly sheared state), as indicated by the black circles.
We also considered the different temperature case of the state at $T=0.267$ and $\dot{\gamma}_{ss}=10^{-3}$, as indicated by the black square.
}
\label{eta}
\end{figure}

As mentioned above, after the quiescent equilibrium state was established, we applied a steady shear flow and an oscillating shear strain.
A steady shear flow is first applied to drive the supercooled model liquid into a non-equilibrium state \cite{yamamoto1_1998,miyazaki_2004,bessel_2007}.
We orient the $x$ and $y$ axes along the flow direction and the velocity gradient direction of the steady shear flow, respectively, as shown in Fig. \ref{visual}.
We denote the shear rate of the steady shear flow as $\dot{\gamma}_{ss}$, where the subscript ``$ss$" indicates ``Steady Shear flow".
Figure \ref{eta} illustrates the shear rate $\dot{\gamma}_{ss}$ dependence of the shear viscosity $\eta$ at various temperatures $T=0.352-0.267$.
The value of $\eta$ decreases with increasing $\dot{\gamma}_{ss}$ as $\eta \sim \dot{\gamma}_{ss}^{-\nu}$ with $\nu$ \hspace{-0.1em}\raisebox{0.4ex}{$<$}\hspace{-0.75em}\raisebox{-.7ex}{$\sim$}\hspace{-0.1em} $1.0$, as demonstrated in previous studies \cite{yamamoto1_1998,miyazaki_2004,bessel_2007}.
As observed in Fig. \ref{eta}, the viscosity $\eta$ displays a good fit to the functional form $\eta = \eta_{s0} /(1+\mu \dot{\gamma}_{ss}^\nu) + \eta_f$ (the viscosity of constitutive Eq. (\ref{conequ})).
Note that this form $\eta = \eta_{s0} /(1+\mu \dot{\gamma}_{ss}^\nu) + \eta_f$ is the same as that proposed for pseudoplastic systems in Ref. \cite{cross_1965}.

In the present study, we mainly focused on two sheared non-equilibrium states, as indicated by the black circles in Fig. \ref{eta}.
One state occurs at $T = 0.306$ and $\dot{\gamma}_{ss} = 10^{-4}$, for which the shear flow is weak and thus the supercooled liquid is nearly in a Newtonian regime, i.e., the supercooled liquid is in the weakly sheared state.
The other state considered is at $T = 0.306$ and $\dot{\gamma}_{ss} = 10^{-2}$, for which the shear flow is so strong that marked shear thinning occurs, i.e., the supercooled liquid is in the strongly sheared state.
We note that a recent study \cite{Chattoraj_2011} discussed the crossover from a Newtonian regime to a non-Newtonian regime (shear thinning regime) for sheared glassy systems in detail.
To ensure that our simulations incorporated a different temperature case, we also considered the state at $T=0.267$ and $\dot{\gamma}_{ss}=10^{-3}$, as indicated by the black square in Fig. \ref{eta}.

\subsection{Oscillating shear strain} \label{sec:2-3}
\begin{figure}
\begin{center}
\includegraphics[scale=1]{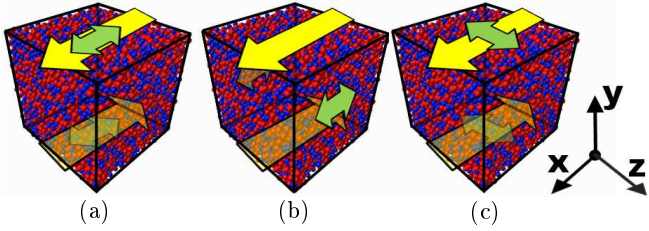}
\end{center}
\vspace*{0mm}
\caption{
Schematic illustration of a steady shear flow (single-headed arrow) and an oscillating shear strain (double-headed arrow).
The steady shear flow is applied as the $xy$ component of the strain tensor $\vec{\gamma}$.
The oscillating shear strain is applied as the $xy$, $xz$, and $yz$ components of $\vec{\gamma}$ in (a), (b), and (c), respectively.
}
\label{visual}
\end{figure}

After the steady sheared state was achieved, we next added an oscillating shear strain to the main drive, as shown in Fig. \ref{visual}.
The oscillating shear strain was applied in a sinusoidal form via the SLLOD algorithm \cite{nonequ}.
We represent the oscillating shear strain as $\delta \gamma^{ij}$ and its shear stress response as $\delta \sigma^{ij}$, where $\gamma^{ij}$ and $\sigma^{ij}$ are the $ij$ components of the strain tensor $\vec{\gamma}$ and the stress tensor $\vec{\sigma}$, respectively, and $ij=xy$, $xz$, and $yz$.
The difference in a quantity from its value in the absence of an oscillating strain is denoted as $\delta$.
Notably, even in the absence of an oscillating strain, $\sigma^{xy}$ has a value $\sigma_{ss} = \eta(\dot{\gamma}_{ss}) \dot{\gamma}_{ss}$ due to the steady shear flow $\dot{\gamma}_{ss}$, and we should therefore calculate $\delta \sigma^{xy}$ as $\delta \sigma^{xy} = \sigma^{xy}- \sigma_{ss}$.
($\delta \sigma^{xz} = \sigma^{xz}$ and $\delta \sigma^{yz} = \sigma^{yz}$.)
Here, we stress that according to the three components $ij=xy$, $xz$, and $yz$, there are three different ways to apply an oscillating shear strain, as shown in Fig. \ref{visual}.
Figs. \ref{visual}(a), (b), and (c) correspond to $ij=xy$, $xz$, and $yz$, respectively.
The strain tensor, $\vec{\gamma} =\nabla \vec{u} + \tr{(\nabla \vec{u})}$, is written as Eqs. (\ref{casexy}), (\ref{casexz}), and (\ref{caseyz}) for $ij=xy$, $xz$, and $yz$, respectively.
\begin{equation}
\begin{aligned}
\vec{\gamma} =
\left(
\begin{array}{ccc}
0 & \dot{\gamma}_{ss} t + \delta \gamma^{xy} & 0 \\
\dot{\gamma}_{ss} t + \delta \gamma^{xy}  & 0 & 0 \\
0  & 0  & 0
\end{array}
\right),
\end{aligned} \label{casexy}
\end{equation}
\begin{equation}
\begin{aligned}
\vec{\gamma} =
\left(
\begin{array}{ccc}
0 & \dot{\gamma}_{ss} t & \delta \gamma^{xz}  \\
\dot{\gamma}_{ss} t  & 0 & 0 \\
\delta \gamma^{xz}  & 0  & 0
\end{array}
\right),
\end{aligned} \label{casexz}
\end{equation}
\begin{equation}
\begin{aligned}
\vec{\gamma} =
\left(
\begin{array}{ccc}
0 & \dot{\gamma}_{ss} t & 0  \\
\dot{\gamma}_{ss} t & 0 & \delta \gamma^{yz} \\
0  & \delta \gamma^{yz}  & 0
\end{array}
\right).
\end{aligned} \label{caseyz}
\end{equation}
In the present situation, in which a steady shear flow with $\dot{\gamma}_{ss}$ is applied, the three stress responses, $\delta \sigma^{xy}$ to $\delta \gamma^{xy}$, $\delta \sigma^{xz}$ to $\delta \gamma^{xz}$, and $\delta \sigma^{yz}$ to $\delta \gamma^{yz}$, are generally different, whereas these three responses are exactly the same in the equilibrium state $\dot{\gamma}_{ss}=0$.

The oscillating shear strain was expressed in a sinusoidal form:
\begin{equation}
\delta \gamma^{ij} = \gamma_0 \sin (\omega t),
\end{equation}
where $\gamma_0$ and $\omega$ are the amplitude and the frequency of the oscillating strain, respectively.
In this study, we set the amplitude $\gamma_0$ to be $\gamma_0 = 0.01-0.2$.
The amplitude $\gamma_0=0.01$ is small enough that the response $\delta \sigma^{ij}$ is linear with respect to $\delta \gamma^{ij}$.
In contrast, if $\gamma_0 > 0.01$, $\delta \sigma^{ij}$ becomes non-linear with respect to $\delta \gamma^{ij}$.

It is beneficial to use the shear moduli $G'^{ij}_{\dot{\gamma}_{ss}}(\omega)$ and $G''^{ij}_{\dot{\gamma}_{ss}}(\omega)$ instead of the full time history $\delta \sigma^{ij}(t)$, where the subscript ``$\dot{\gamma}_{ss}$" denotes a value in the sheared non-equilibrium state.
The values $G'^{ij}_{\dot{\gamma}_{ss}}(\omega)$ and $G''^{ij}_{\dot{\gamma}_{ss}}(\omega)$ are the storage modulus for the elasticity and the loss modulus for the viscosity, respectively, and are often used to measure the viscoelastic properties of the materials \cite{miyazaki_2006,brader_2010,wyss_2007,yasuda_2011}.
We can calculate $G'^{ij}_{\dot{\gamma}_{ss}}(\omega)$ and $G''^{ij}_{\dot{\gamma}_{ss}}(\omega)$ as the Fourier transformations of $\delta \sigma^{ij}(t)$:
\begin{equation}
\begin{aligned}
G'^{ij}_{\dot{\gamma}_{ss}}(\omega) &= \frac{\omega}{\pi} \int_{-\pi/\omega}^{\pi/\omega} \frac{\delta \sigma^{ij}(t)}{\gamma_0} \sin(\omega t) dt, \\
G''^{ij}_{\dot{\gamma}_{ss}}(\omega) &= \frac{\omega}{\pi} \int_{-\pi/\omega}^{\pi/\omega} \frac{\delta \sigma^{ij}(t)}{\gamma_0} \cos(\omega t) dt.
\end{aligned}
\end{equation}
In this case, the time history $\delta \sigma^{ij}(t)$ can be expressed as $\delta \sigma^{ij}(t) = G'^{ij}_{\dot{\gamma}_{ss}}(\omega) \gamma_0 \sin(\omega t) + G''^{ij}_{\dot{\gamma}_{ss}}(\omega) \gamma_0 \cos(\omega t)$.
The shear moduli $G'^{ij}_{\dot{\gamma}_{ss}}(\omega)$ and $G''^{ij}_{\dot{\gamma}_{ss}}(\omega)$ depend on the three quantities $\omega$, $\gamma_0$, and $\dot{\gamma}_{ss}$.
If the amplitude $\gamma_0$ of the oscillating strain is small and the steady shear flow $\dot{\gamma}_{ss}$ is weak, then $G'^{ij}_{\dot{\gamma}_{ss}}(\omega)$ and $G''^{ij}_{\dot{\gamma}_{ss}}(\omega)$ reduce to the linear shear moduli and depend only on the frequency $\omega$ of the oscillating strain.
However, when the amplitude $\gamma_0$ becomes large or the steady shear flow $\dot{\gamma}_{ss}$ becomes strong, then significant non-linearity arises, so $G'^{ij}_{\dot{\gamma}_{ss}}(\omega)$ and $G''^{ij}_{\dot{\gamma}_{ss}}(\omega)$ become the non-linear shear moduli \cite{miyazaki_2006,brader_2010,wyss_2007,yasuda_2011}, which depend not only on the frequency $\omega$ but also the amplitude $\gamma_0$ or the steady shear rate $\dot{\gamma}_{ss}$.
We should note that when the responses are non-linear with respect to the oscillating strain, there are higher harmonic contributions to the responses \cite{brader_2010,wyss_2007,yasuda_2011}.
In the present study, we considered only the first harmonic contribution, i.e., the shear moduli $G'^{ij}_{\dot{\gamma}_{ss}}(\omega)$ and $G''^{ij}_{\dot{\gamma}_{ss}}(\omega)$, which were verified to exert a dominant effect compared with the higher harmonic contributions.

\section{Result I: Mechanical responses} \label{sec:3}
In this section, the results of mechanical responses are discussed.
We present the shear moduli $G'^{ij}_{\dot{\gamma}_{ss}}$ and $G''^{ij}_{\dot{\gamma}_{ss}}$ obtained from MD simulations.
The simulation cases are summarized in Table \ref{tab:1}.
As we mentioned in Sect.~\ref{sec:2-2}, we primarily focused on two sheared states with the temperature $T=0.306$: the weakly sheared state $\dot{\gamma}_{ss}=10^{-4}$ and the strongly sheared state $\dot{\gamma}_{ss}=10^{-2}$.
In addition to these two states, we considered another state with $T=0.267$ as a case involving a different temperature.
We also present the constitutive equation that captures the key features of the simulation results.

\subsection{Mechanical responses in the weakly sheared state} \label{sec:3-1}
\begin{figure}
\begin{center}
\includegraphics[scale=1]{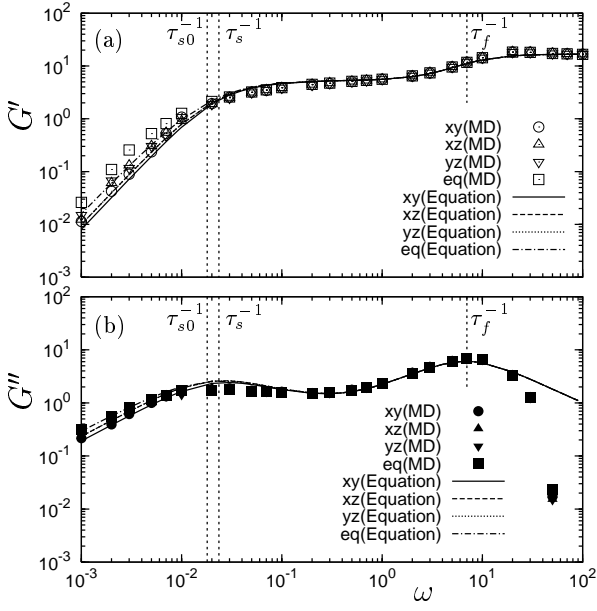}
\end{center}
\vspace*{0mm}
\caption{
The shear moduli (a) $G'^{ij}_{\dot{\gamma}_{ss}}$ and (b) $G''^{ij}_{\dot{\gamma}_{ss}}$ versus the frequency $\omega$ at $T=0.306$ and $\dot{\gamma}_{ss}=10^{-4}$ (weakly sheared case), with the small amplitude of $\gamma_0=0.01$.
We also show $G'_{\text{eq}}$ and $G''_{\text{eq}}$ in the equilibrium state $\dot{\gamma}_{ss}=0$.
The symbols and lines represent the results of MD simulations and constitutive equation (\ref{conequ}), respectively.
The labels ``$xy$", ``$xz$", and ``$yz$" denote the component $ij=xy$, $xz$, and $yz$, respectively, and the label ``eq" denotes the equilibrium state.
We indicate the two time scales of the constitutive equation (\ref{conequ}): $\tau_{s}$ and $\tau_f$ in the sheared state and $\tau_{s0}$ and $\tau_f$ in the equilibrium state.
}
\label{result1a}
\end{figure}

\begin{figure}
\begin{center}
\includegraphics[scale=1]{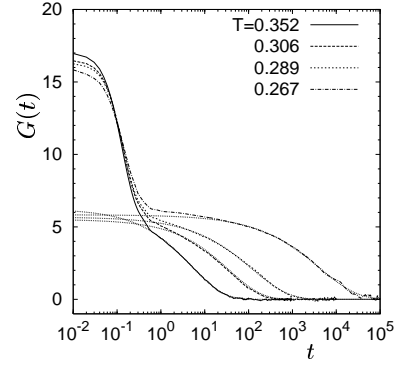}
\end{center}
\vspace*{0mm}
\caption{
The shear stress correlation function $G(t)$ in the equilibrium state at various temperatures.
At long times, the function $G(t)$ can be well fitted by the stretch exponential form $G_{\alpha} \exp \left( -(t/\tau_{\alpha})^\psi \right)$ as indicated by dotted lines \cite{furukawa_2011}, where the value $\tau_\alpha$ is the $\alpha$-relaxation time, and the value $G_\alpha$ is the plateau modulus.
}
\label{scfpa}
\end{figure}

We first present the results of the mechanical responses in the weakly sheared state $T=0.306$ and $\dot{\gamma}_{ss}=10^{-4}$.
Figure \ref{result1a} illustrates the shear moduli $G'^{ij}_{\dot{\gamma}_{ss}}$ and $G''^{ij}_{\dot{\gamma}_{ss}}$ at the small amplitude $\gamma_0=0.01$ of the oscillating shear strain.
The value $\gamma_0=0.01$ is small enough that the mechanical response is linear with respect to the oscillating strain.
In the same figure, we also show the values of the shear moduli $G'_{\text{eq}}$ and $G''_{\text{eq}}$ in the equilibrium state to clarify the effects caused by the steady shear flow $\dot{\gamma}_{ss}$.
As can be seen from Fig. \ref{result1a}, the shear moduli $G'_{\text{eq}}$ and $G''_{\text{eq}}$ demonstrate the typical dependence on frequency $\omega$ of the Maxwell model with two time scales \cite{polymetric,transport}.
In Fig. \ref{result1a}, we indicate these two time scales as $\tau_{s0}$ (slower time) and $\tau_f$ (faster time).
As is well known, the stress correlation functions of supercooled liquids exhibit two-step relaxation \cite{Debenedetti_2001,kim1_2010} (see also Fig. \ref{scfpa}).
The slower relaxation is called $\alpha$-relaxation, and its relaxation time is thus known as the $\alpha$-relaxation time $\tau_\alpha$.
The faster relaxation results from the thermal vibrations of the particles, and its time scale is known as the Einstein period $\tau_E$ ($\omega_E=2\pi/\tau_E$ is the Einstein frequency).
The two time scales $\tau_{s0}$ and $\tau_f$ are equivalent to $\tau_\alpha$ and $\tau_E$, respectively ($\tau_{s0} \sim \tau_\alpha \gg \tau_f \sim \tau_E$).
Here, we note that at approximately the slower time scale $\tau_{s0} \sim \tau_\alpha$, supercooled liquids exhibit the crossover from liquid-like behavior to solid-like behavior.
At low frequencies $\omega$, $G''_{\text{eq}}$ is larger than $G'_{\text{eq}}$, which results in liquid-like behavior.
As $\omega$ becomes large, $G'_{\text{eq}}$ becomes larger than $G''_{\text{eq}}$, which results in solid-like behavior.
At $\omega \sim \tau_{s0}^{-1}$, $G'_{\text{eq}} \simeq G''_{\text{eq}}$, i.e., the crossover occurs.
In addition, as we will explain in detail in Sect.~\ref{sec:3-3}, $G'^{ij}_{\dot{\gamma}_{ss}}$ and $G''^{ij}_{\dot{\gamma}_{ss}}$ can also be well described by the two-mode Maxwell model as in Eq. (\ref{conequ}); therefore, the mechanical properties of the supercooled liquid can be characterized by two time scales not only in the equilibrium state but also in the sheared non-equilibrium state.
In Fig. \ref{result1a}, two time scales in the sheared state are indicated as $\tau_{s}$ (slower time) and $\tau_{f}$ (faster time).

As in Fig. \ref{result1a}, although we can recognize the effects due to the steady shear flow at the very low frequencies of $\omega < \tau_s^{-1}$, these effects are very small.
In the whole frequency $\omega$ region except for $\omega < \tau_s^{-1}$, all $G'^{ij}_{\dot{\gamma}_{ss}}$ and $G''^{ij}_{\dot{\gamma}_{ss}}$ values almost coincide with $G'_{\text{eq}}$ and $G''_{\text{eq}}$.
The slower time scale $\tau_s$ (equivalent to $\tau_\alpha$) is also close to the equilibrium time scale of $\tau_{s0}$, although $\tau_s$ is a little shorter than $\tau_{s0}$.
The faster time scale (equivalent to $\tau_E$) is unchanged at $\tau_f$.
Thus, the weak steady shear flow with $\dot{\gamma}_{ss}=10^{-4}$ produces only small effects on the mechanical responses in the very low frequency region $\omega <\tau_s^{-1}$.

\begin{figure}
\begin{center}
\includegraphics[scale=1]{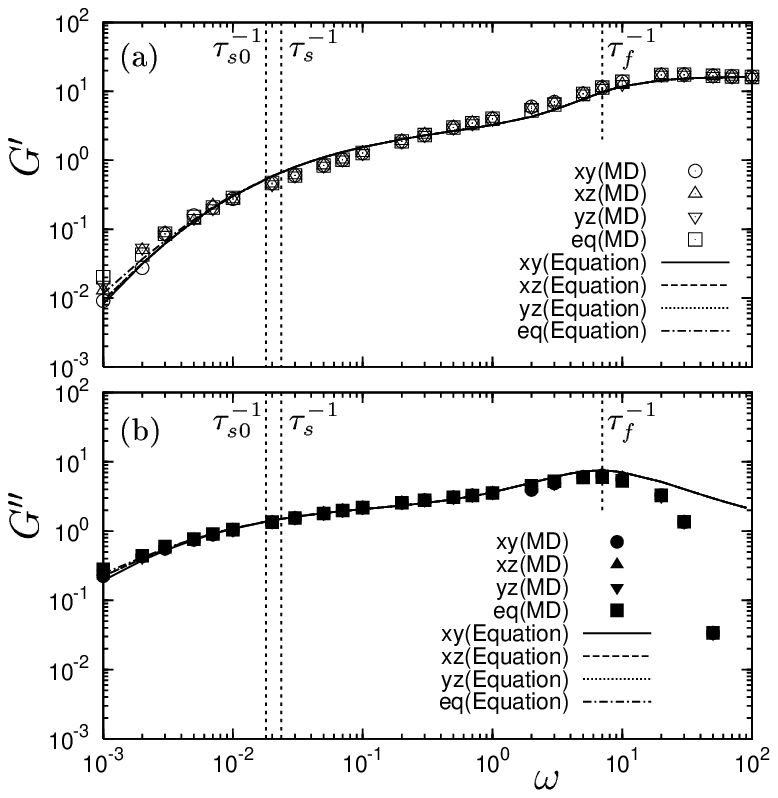}
\end{center}
\vspace*{0mm}
\caption{
The shear moduli (a) $G'^{ij}_{\dot{\gamma}_{ss}}$ and (b) $G''^{ij}_{\dot{\gamma}_{ss}}$ versus the frequency $\omega$ at $T=0.306$ and $\dot{\gamma}_{ss}=10^{-4}$ (weakly sheared case), with the large amplitude of $\gamma_0=0.1$.
See also the caption of Fig. \ref{result1a}.
}
\label{result1b}
\end{figure}

Furthermore, in Fig. \ref{result1b}, we demonstrate $G'^{ij}_{\dot{\gamma}_{ss}}$ and $G''^{ij}_{\dot{\gamma}_{ss}}$ at the large amplitude $\gamma_0=0.1$ of the oscillating shear strain. 
The value $\gamma_0=0.1$ is large; thus, the mechanical response is non-linear with respect to the oscillating strain.
Comparing Fig. \ref{result1b} ($\gamma_0=0.1$) with Fig. \ref{result1a} ($\gamma_0=0.01$), we can observe that the storage modulus $G'^{ij}_{\dot{\gamma}_{ss}}$ decreases for a greater $\gamma_0$, and the loss modulus $G''^{ij}_{\dot{\gamma}_{ss}}$ becomes larger relatively.
This observation implies that the larger oscillating strain makes the supercooled liquid more liquid-like. Such non-linear viscoelasticity has also been observed in soft materials \cite{miyazaki_2006,wyss_2007}, dense colloidal suspensions \cite{brader_2010}, and supercooled polymer melts \cite{yasuda_2011}.
In addition, it is also notable that the effects of the steady shear flow $\dot{\gamma}_{ss}$ become smaller for a greater $\gamma_0=0.1$, as the oscillating shear strain becomes relatively strong compared with the steady shear flow.

\subsection{Mechanical responses in the strongly sheared state} \label{sec:3-2}
\begin{figure}
\begin{center}
\includegraphics[scale=1]{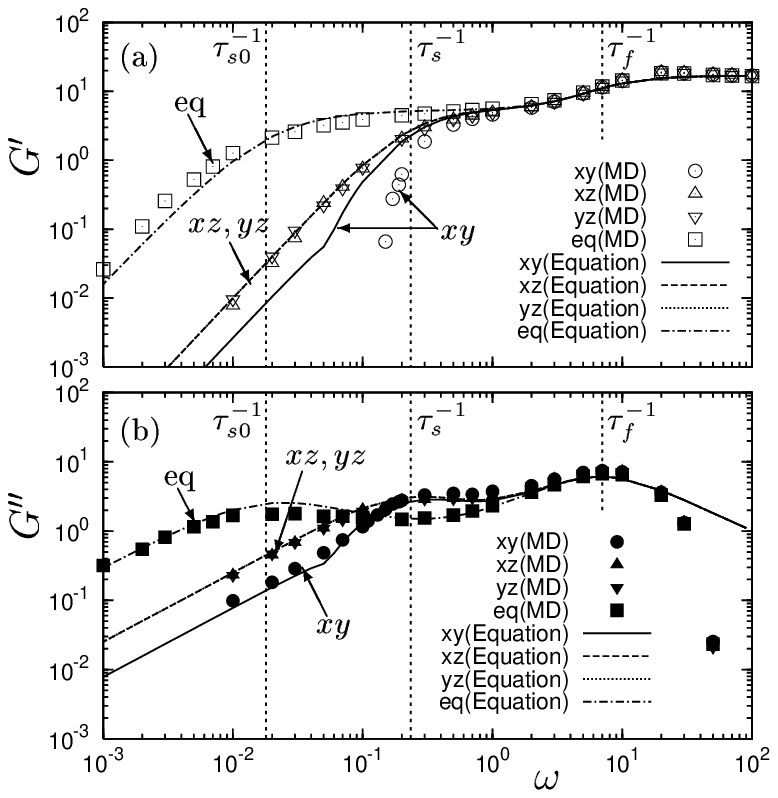}
\end{center}
\vspace*{0mm}
\caption{
The shear moduli (a) $G'^{ij}_{\dot{\gamma}_{ss}}$ and (b) $G''^{ij}_{\dot{\gamma}_{ss}}$ versus the frequency $\omega$ at $T=0.306$ and $\dot{\gamma}_{ss}=10^{-2}$ (strongly sheared case), with the small amplitude of $\gamma_0=0.01$.
See also the caption of Fig. \ref{result1a}.
}
\label{result2a}
\end{figure}

\begin{figure}
\begin{center}
\includegraphics[scale=1]{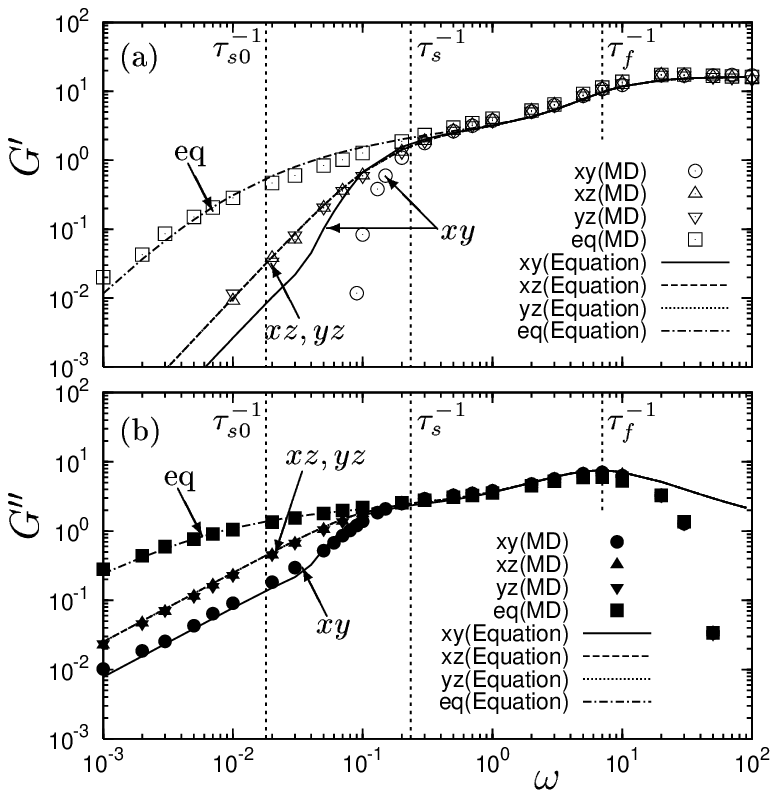}
\end{center}
\vspace*{0mm}
\caption{
The shear moduli (a) $G'^{ij}_{\dot{\gamma}_{ss}}$ and (b) $G''^{ij}_{\dot{\gamma}_{ss}}$ versus the frequency $\omega$ at $T=0.306$ and $\dot{\gamma}_{ss}=10^{-2}$ (strongly sheared case), with the large amplitude of $\gamma_0=0.1$.
See also the caption of Fig. \ref{result1a}.
}
\label{result2b}
\end{figure}

Figures \ref{result2a} and \ref{result2b} illustrate the results for $G'^{ij}_{\dot{\gamma}_{ss}}$ and $G''^{ij}_{\dot{\gamma}_{ss}}$ in the strongly sheared state $T=0.306$ and $\dot{\gamma}_{ss}=10^{-2}$.
The amplitude $\gamma_0$ of the oscillating shear strain is $\gamma_0=0.01$ (linear regime) in Fig. \ref{result2a} and $\gamma_0=0.1$ (non-linear regime) in Fig. \ref{result2b}.
As in Figs. \ref{result2a} and \ref{result2b}, we can easily recognize the effects resulting from the steady shear flow at low frequencies $\omega < \tau_s^{-1}$ for all $G'^{ij}_{\dot{\gamma}_{ss}}$ and $G''^{ij}_{\dot{\gamma}_{ss}}$, whereas at high frequencies $\omega> \tau_f^{-1}$, these effects are not observed, i.e., all $G'^{ij}_{\dot{\gamma}_{ss}}$ and $G''^{ij}_{\dot{\gamma}_{ss}}$ values coincide with $G'_{\text{eq}}$ and $G''_{\text{eq}}$.
Notice that due to the steady shear flow, the slower time scale (equivalent to $\tau_\alpha$) becomes dramatically shortened from $\tau_{s0}$ to $\tau_s$, whereas the faster time scale (equivalent to $\tau_E$) is unchanged at $\tau_f$.
The strong shear flow with $\dot{\gamma}_{ss}=10^{-2}$ influences the mechanical responses to a much greater extent than is observed for the weak shear flow with $\dot{\gamma}_{ss}=10^{-4}$.

From Figs. \ref{result2a} and \ref{result2b}, we obtain two remarkable results for both the linear and the non-linear responses (both $\gamma_0=0.01$ and $\gamma_0=0.1$).
First, two components, $xz$ and $yz$, of $G'^{ij}_{\dot{\gamma}_{ss}}$ and $G''^{ij}_{\dot{\gamma}_{ss}}$ coincide with each other surprisingly well, even at low frequencies $\omega < \tau_s^{-1}$ (refer to the upper and lower triangles in Figs. \ref{result2a} and \ref{result2b}).
Despite the strong steady shear flow, two of the stress responses, $\delta \sigma_{xz}$ to $\delta \gamma_{xz}$ and $\delta \sigma_{yz}$ to $\delta \gamma_{yz}$, are the same.
This result demonstrates the isotropic aspect of this system.
Second, the behavior of the $xy$ component (refer to the circles in Figs. \ref{result2a} and \ref{result2b}) is quite different from those of the $xz$ and $yz$ components.
The $xy$ component is smaller than either the $xz$ or $yz$ component at low $\omega$.
In addition, as $\omega$ decreases, the storage modulus $G'^{xy}_{\dot{\gamma}_{ss}}$ decreases much more rapidly than $G'^{xz}_{\dot{\gamma}_{ss}}$ and $G'^{yz}_{\dot{\gamma}_{ss}}$.
At low $\omega$, $G'^{xy}_{\dot{\gamma}_{ss}}$ takes on negative values (data are not shown in Figs. \ref{result2a} and \ref{result2b}).
Thus, due to the steady shear flow, the mechanical properties of the $xy$ component are notably different from those of the other components.
This result demonstrates the anisotropic responses of this system.
We will discuss the origin of the anisotropic responses, i.e., the difference between the $xy$ component and the $xz$ and $yz$ components, in Sect.~\ref{sec:3-4}.
Similar behaviors of $G'^{xy}_{\dot{\gamma}_{ss}}$ and $G''^{xy}_{\dot{\gamma}_{ss}}$ have been previously observed in polymer solutions \cite{vermant_1998,somma_2007}.

\begin{figure}
\begin{center}
\includegraphics[scale=1]{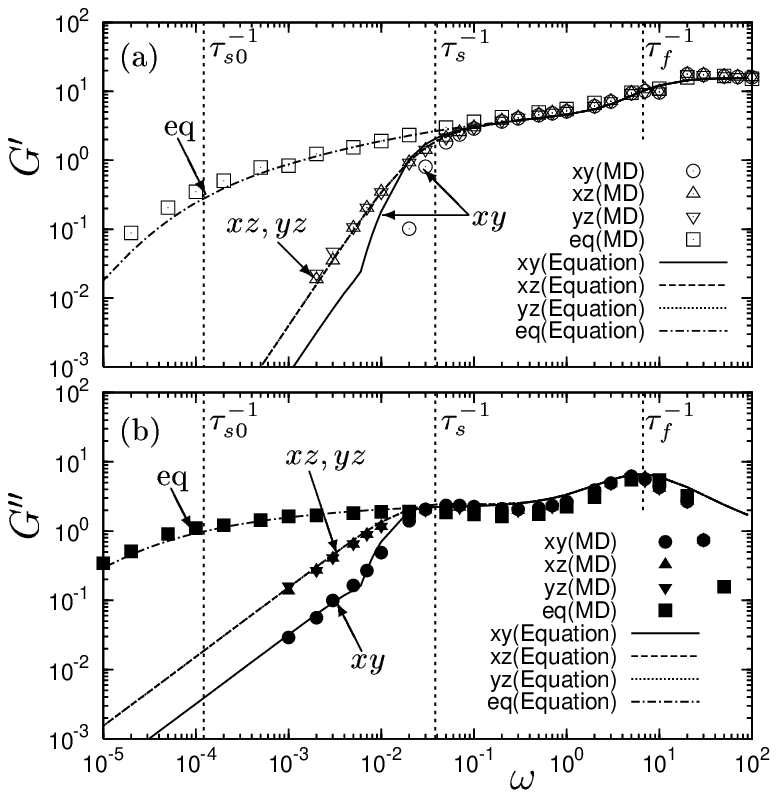}
\end{center}
\vspace*{0mm}
\caption{
The shear moduli (a) $G'^{ij}_{\dot{\gamma}_{ss}}$ and (b) $G''^{ij}_{\dot{\gamma}_{ss}}$ versus the frequency $\omega$ at $T=0.267$ and $\dot{\gamma}_{ss}=10^{-3}$ (different temperature case), with the small amplitude of $\gamma_0=0.05$.
See also the caption of Fig. \ref{result1a}.
}
\label{result3a}
\end{figure}

\begin{figure}
\begin{center}
\includegraphics[scale=1]{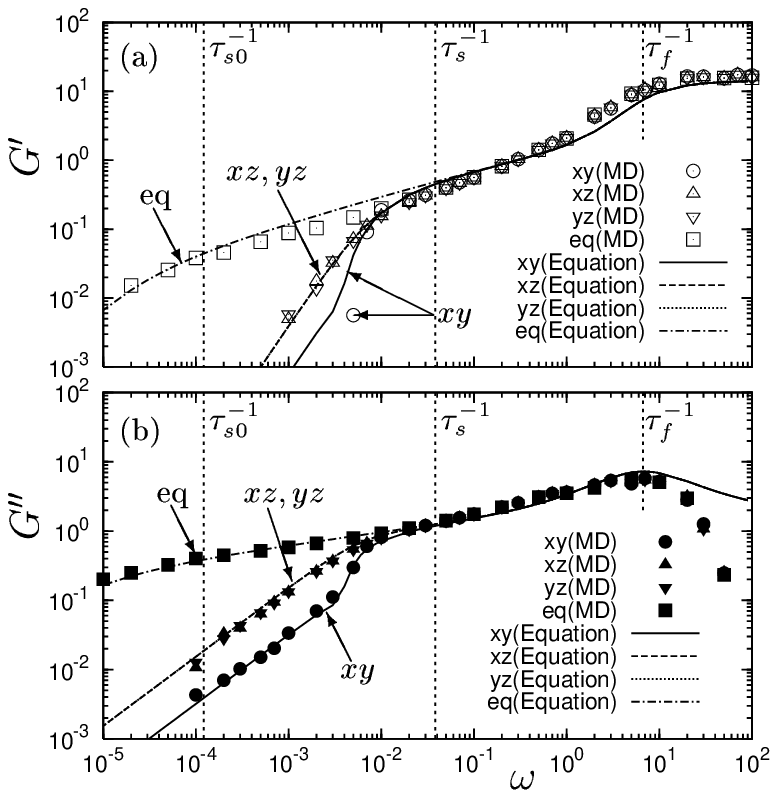}
\end{center}
\vspace*{0mm}
\caption{
The shear moduli (a) $G'^{ij}_{\dot{\gamma}_{ss}}$ and (b) $G''^{ij}_{\dot{\gamma}_{ss}}$ versus the frequency $\omega$ at $T=0.267$ and $\dot{\gamma}_{ss}=10^{-3}$ (different temperature case), with the large amplitude of $\gamma_0=0.2$.
See also the caption of Fig. \ref{result1a}.
}
\label{result3b}
\end{figure}

In Figs. \ref{result3a} and \ref{result3b}, we present $G'^{ij}_{\dot{\gamma}_{ss}}$ and $G''^{ij}_{\dot{\gamma}_{ss}}$ at the different temperature case of $T=0.267$ and $\dot{\gamma}_{ss}=10^{-3}$, for which the supercooled liquid is also in the strongly sheared state.
Figures \ref{result3a} and \ref{result3b} show the results for the different amplitudes $\gamma_0=0.05$ and $0.2$, respectively.
We can confirm that the same observations from Figs. \ref{result2a} and \ref{result2b} are also obtained in the scenario displayed in Figs. \ref{result3a} and \ref{result3b}.

\subsection{Constitutive equation} \label{sec:3-3}
We attempted to construct a constitutive equation describing our simulation results and obtained the following two-mode Maxwell model equation:
\begin{equation}
\begin{aligned}
& {\sigma}^{ij}_{s} + {\tau}_{s}(\dot{\vec{\gamma}}) \frac{d {\sigma}^{ij}_{s}}{dt} = \eta_{s}(\dot{\vec{\gamma}}) \dot{\vec{\gamma}}^{ij}, \\
& {\sigma}^{ij}_{f} + {\tau}_{f} \frac{d {\sigma}^{ij}_{f}}{dt} = \eta_{f} \dot{\vec{\gamma}}^{ij},
\label{conequ}
\end{aligned}
\end{equation}
with
\begin{equation}
\begin{aligned}
& {\tau}_{s}(\dot{\vec{\gamma}}) = \frac{\tau_{s0}}{1 + \mu  (\dot{\gamma}^{xy 2}+\dot{\gamma}^{xz 2}+\dot{\gamma}^{yz 2})^{\nu/2} }, \\
& {\eta}_{s}(\dot{\vec{\gamma}}) = \frac{\eta_{s0}}{1 + \mu  (\dot{\gamma}^{xy 2}+\dot{\gamma}^{xz 2}+\dot{\gamma}^{yz 2})^{\nu/2} }, \label{param}
\end{aligned}
\end{equation}
where $ij=xy,\ xz$, or $yz$.
The model equation consists of a slower component and a faster component, which are denoted by the subscripts ``$s$'' and ``$f$'', respectively.
The stress tensor ${\sigma}^{ij}$ is written as ${\sigma}^{ij} = {\sigma}^{ij}_{s} + {\sigma}^{ij}_{f}$.
As in Figs. \ref{result1a} or \ref{result2a}, the (linear) shear moduli $G'_{\text{eq}}$ and $G''_{\text{eq}}$ in the equilibrium state demonstrate the typical frequency $\omega$ dependence of the Maxwell model with two characteristic times \cite{polymetric,transport}; therefore, we considered the two-mode Maxwell model equation.
In this model, the shear viscosity $\eta$ and the instantaneous shear modulus $G_\infty$ are described as $\eta = \eta_{s} + \eta_{f}$ and $G_\infty = G_s + G_f = \eta_s / \tau_s + \eta_f / \tau_f$, respectively.
The values $\eta_s$ and $G_s = \eta_s/\tau_s$ are slower components of $\eta$ and $G_\infty$, whereas the values $\eta_f$ and $G_f = \eta_f/\tau_f$ are faster components.
As we mentioned in Sect.~\ref{sec:3-1}, the two time scales $\tau_{s}$ and $\tau_{f}$ are considered to be equivalent to the $\alpha$-relaxation time $\tau_\alpha$ and the Einstein period $\tau_E$, respectively.
Therefore, we naturally assumed that the values $\tau_f$ and $\eta_f$ characterizing the faster component are constant and unaffected by the applied shear flows, whereas the slower components $\tau_s$ and $\eta_s$ do depend on the applied shear flows.
We set $\tau_s$ and $\eta_s$ as functions of only the total strength of the shear rate $(\dot{\gamma}^{xy 2}+\dot{\gamma}^{xy 2}+\dot{\gamma}^{xy 2})^{1/2}$ as in Eq. (\ref{param}), which reflects the isotropic feature observed in MD simulations, i.e., that the shear responses of the $xz$ and $yz$ components coincide with each other in Figs. \ref{result2a} and \ref{result2b}.
In addition, we assumed $\tau_s \sim \eta_s$ and set the same functional form for $\tau_s$ and $\eta_s$ as in Eq. (\ref{param}).
The functional form simply arises from the shear-thinning behavior shown in Fig. \ref{eta}.
In fact, when we consider a steady shear flow with $\dot{\gamma}_{ss}$, the shear viscosity $\eta$ is described as $\eta = \eta_s + \eta_f = \eta_{s0} /(1+\mu \dot{\gamma}_{ss}^\nu) + \eta_f$, which is precisely the shear-thinning form.
As we mentioned previously, this functional form of $\eta$ is the same as that proposed for pseudoplastic systems in Ref. \cite{cross_1965}.

\begin{figure}
\begin{center}
\includegraphics[scale=0.88]{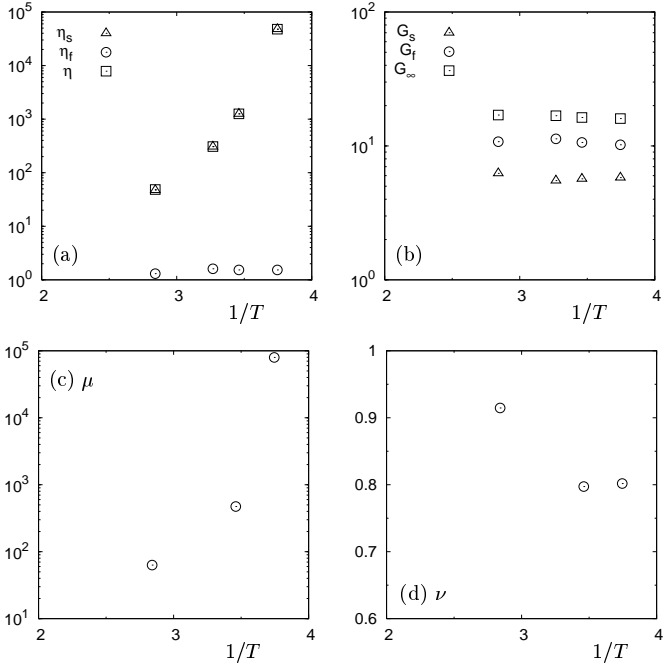}
\end{center}
\vspace*{0mm}
\caption{
The temperature dependences of the parameters (a) $\eta_s$ and $\eta_f$, (b) $G_s$ and $G_f$, (c) $\mu$, and (d) $\nu$.
We also plot the values of the total viscosity $\eta=\eta_s + \eta_f$ in (a) and the total modulus $G_\infty = G_s + G_f$ in (b).
}
\label{parametersf}
\end{figure}

Together, the constitutive Eq. (\ref{conequ}) and Eq. (\ref{param}) have six parameters, $\eta_{s0},\ \eta_f,\ G_{s0}=\eta_{s0}/\tau_{s0},\ G_f=\eta_f/\tau_f,\ \mu$, and $\nu$.
Four of these parameters, $\eta_{s0}$, $\eta_{f}$, $G_{s0}$, and $G_f$, characterize the linear mechanical responses in the equilibrium state.
We can determine these four parameters from the shear stress correlation function $G(t)$, defined as
\begin{equation}
G(t) = \frac{V}{T}{\left< \delta \sigma(t) \delta \sigma(0) \right>_{\text{eq}}},
\end{equation}
where $\delta \sigma$ represents the shear stress fluctuations and $\left< \right>_\text{eq}$ denotes the ensemble average in the equilibrium state.
The shear viscosity $\eta$ and the instantaneous shear modulus $G_\infty$ are related to the function $G(t)$ through $\eta = \int_0^\infty G(t) dt$ and $G_\infty = G(t=0)$ \cite{simpleliquid}.
In addition, as is well demonstrated in Ref. \cite{furukawa_2011} and Fig. \ref{scfpa}, the slower relaxation of $G(t)$ can be well fitted by the stretch exponential form $G_{\alpha} \exp \left( -(t/\tau_{\alpha})^\psi \right)$, where the value $G_\alpha$ is known as the plateau modulus \cite{yoshino_2010,szamel_2011}.
We therefore determined $\eta_{s0}$ and $G_{s0}$ (the slower components of the shear viscosity and modulus) as
\begin{equation}
\begin{aligned}
& \eta_{s0} = \int_{0}^\infty G_{\alpha} \exp \left( -({t}/{\tau_{\alpha}})^\psi \right) dt, \\
& G_{s0} =G_\alpha,
\end{aligned}
\end{equation}
and the values of $\eta_f$ and $G_f$ (the faster components) were determined as
\begin{equation}
\begin{aligned}
& \eta_f=\eta-\eta_{s0}=\int_0^\infty G(t)-G_{\alpha} \exp \left( -(t/\tau_{\alpha})^\psi \right)dt, \\
& G_f=G_\infty-G_{s0}=G(t=0) - G_\alpha.
\end{aligned}
\end{equation}
Figure \ref{parametersf} shows the temperature $T$ dependences of $\eta_{s0}$, $\eta_f$, $G_{s0}$, and $G_f$.
The slower component $\eta_{s0}$ increases dramatically with decreasing temperature $T$ \cite{Debenedetti_2001,kim1_2010}.
The faster component $\eta_f$ is much (several orders of magnitude) smaller than $\eta_{s0}$, and the shear viscosity $\eta$ is almost the same as the slower component $\eta_{s0}$.
On the other hand, the values $G_{s0}$ and $G_f$ are nearly constant with respect to $T$, as was previously observed in Ref. \cite{furukawa_2011}.

\begin{figure}
\begin{center}
\includegraphics[scale=1]{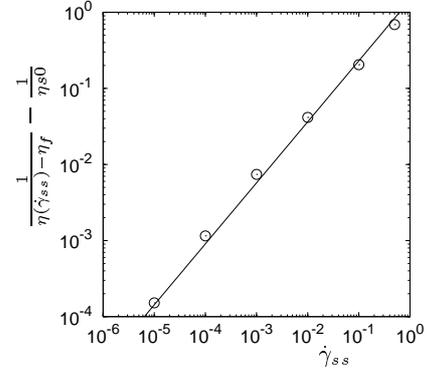}
\end{center}
\vspace*{0mm}
\caption{
The value $1/(\eta-\eta_{f})-1/\eta_{s0}$ versus the shear rate $\dot{\gamma}_{ss}$.
The temperature is $T=0.267$.
The straight line is fitted by least-squares fit.
}
\label{scfpb}
\end{figure}

The remaining two parameters, $\mu$ and $\nu$, characterize the non-linearity resulting from the driving shear flow and can be determined from fitting the shear viscosity $\eta = \eta_{s0} /(1+\mu \dot{\gamma}_{ss}^\nu) + \eta_f$ of the model equation to the simulation data $\eta(\dot{\gamma}_{ss})$ shown in Fig. \ref{eta}.
The functional form $\eta = \eta_{s0} /(1+\mu \dot{\gamma}_{ss}^\nu) + \eta_f$ is transformed to
\begin{equation}
\frac{1}{\eta-\eta_f} - \frac{1}{\eta_{s0}} = \left(\frac{\mu}{\eta_{s0}} \right) \dot{\gamma}_{ss}^\nu,
\end{equation}
and thus, we fitted the function $(\mu/\eta_{s0}) \dot{\gamma}_{ss}^\nu$ to the data $1/(\eta(\dot{\gamma}_{ss})-\eta_f) - 1/\eta_{s0}$, as in Ref. \cite{cross_1965}.
As shown in Fig. \ref{scfpb}, the straight line is well fitted in the log-log plot, for which we performed the least-squares fit.
The slope and intercept of the fitted line then correspond to the values $\nu$ and $\mu/\eta_{s0}$, respectively.
As can be observed in Fig. \ref{eta}, the function $\eta = \eta_{s0} /(1+\mu \dot{\gamma}_{ss}^\nu) + \eta_f$ with the obtained values of $\mu$ and $\nu$ demonstrate a good fit to the simulation data $\eta(\dot{\gamma}_{ss})$.
Figure \ref{parametersf} also illustrates the temperature $T$ dependences of $\mu$ and $\nu$.
The value $\mu$ increases drastically with decreasing $T$ in a similar way as the viscosity $\eta$, whereas $\nu$ is insensitive to $T$ and takes values between $0.8$ and $1.0$ \cite{yamamoto1_1998,miyazaki_2004,bessel_2007}.

We calculated the shear moduli $G'^{ij}_{\dot{\gamma}_{ss}}$ and $G''^{ij}_{\dot{\gamma}_{ss}}$ using the constitutive equation (\ref{conequ}) with the parameters presented in Fig. \ref{parametersf}.
Figures \ref{result1a} and \ref{result1b} also show the results of $G'^{ij}_{\dot{\gamma}_{ss}}$ and $G''^{ij}_{\dot{\gamma}_{ss}}$ obtained from Eq. (\ref{conequ}) in the weakly sheared state of $T=0.306$ and $\dot{\gamma}_{ss}=10^{-4}$.
By comparing the lines (constitutive equation) with the symbols (MD simulation), we can clearly observe that Eq. (\ref{conequ}) reflects the results of the MD simulations quite well in both the equilibrium state and the weakly sheared state.
Notice that the constitutive equation can account for not only the linear responses (the amplitude $\gamma_0=0.01$ in Fig. \ref{result1a}) but also the non-linear responses ($\gamma_0=0.1$ in Fig. \ref{result1b}).
In addition, we show $G'^{ij}_{\dot{\gamma}_{ss}}$ and $G''^{ij}_{\dot{\gamma}_{ss}}$ of Eq. (\ref{conequ}) in the strongly sheared state $T=0.306$ and $\dot{\gamma}_{ss}=10^{-2}$ in Figs. \ref{result2a} and \ref{result2b}.
These results demonstrate that the constitutive equation also functions surprisingly well even in the strongly sheared state, except for the storage modulus $G'^{xy}_{\dot{\gamma}_{ss}}$ at low frequencies $\omega$.
The modulus $G'^{xy}_{\dot{\gamma}_{ss}}$ takes on negative values at low $\omega$, as we mentioned previously in Sect.~\ref{sec:3-2}, and Eq. (\ref{conequ}) cannot account for this negative storage modulus.
At this stage, we do not understand the mechanism and the origin of the negative values in question for $G'^{xy}_{\dot{\gamma}_{ss}}$, and this topic will be a subject of future work.
Once the mechanisms underlying these values are more fully elucidated, we will be able to propose certain modifications of the constitutive equation to account for this negative modulus.

Furthermore, we verified the validity of the constitutive equation in a case involving a different temperature $T=0.267$ and $\dot{\gamma}_{ss}=10^{-3}$, as shown in Figs. \ref{result3a} and \ref{result3b}.
In Figs. \ref{result3a} and \ref{result3b}, we once again observe that the constitutive equation is valid except for the storage modulus $G'^{xy}_{\dot{\gamma}_{ss}}$ at low frequencies $\omega$.
We stress that all six parameters of the constitutive equation have physical significance and can only be completely determined by the mechanical properties ($G(t)$ and $\eta(\dot{\gamma}_{ss})$) in the equilibrium situation and the steady sheared situation.
Using these six parameters, we can accurately predict mechanical properties in more general situations, e.g., under two different shear strains (the steady shear flow and the oscillating strain).
The constitutive Eq. (\ref{conequ}) is much simpler than other equations obtained for typical complex fluids such as polymer solutions \cite{polymetric,transport}, and interestingly, even in the strongly sheared state, the mechanical responses can be well fit by this simple constitutive equation.

\subsection{Origin of anisotropic mechanical responses} \label{sec:3-4}
As demonstrated in Figs. \ref{result2a} and \ref{result2b} (and also in Figs. \ref{result3a} and \ref{result3b}), $G'^{xy}_{\dot{\gamma}_{ss}}$ and $G''^{xy}_{\dot{\gamma}_{ss}}$ differ from the other $xz$ and $yz$ components at low frequencies $\omega$ in the strongly sheared state.
We can understand the origin of this difference through the examination of the constitutive equation (\ref{conequ}).
At low $\omega$, the slower component ${\sigma}_s^{ij}$ is dominant compared with the faster component $\sigma_f^{ij}$; thus, in the following analysis, we consider only ${\sigma}_s^{ij}$.
To reveal the effects due to the steady shear flow, we analyzed the constitutive equation under the condition $\gamma_0 \omega \ll \dot{\gamma}_{ss}$.
This condition, $\gamma_0 \omega \ll \dot{\gamma}_{ss}$, indicates that the shear rate of the steady shear flow is much larger than that of the oscillating strain, which holds true for low $\omega$.
Under this condition, the following two equations were obtained: Eq. (\ref{parallel}) for the $xy$ component and Eq. (\ref{orthogonal}) for the $xz$ and $yz$ components.
\begin{equation}
\begin{aligned}
\delta {\sigma}^{xy}_s + {\tau}_{s}\frac{d \delta{\sigma}^{xy}_s}{dt} = \eta_{s} \delta {\dot{\gamma}^{xy}} - \frac{\mu \nu \eta_{s}}{(1+\mu \dot{\gamma}_{ss}^\nu)} \dot{\gamma}_{ss}^{\nu} \delta {\dot{\gamma}}^{xy},
\label{parallel}
\end{aligned}
\end{equation}
\begin{equation}
\begin{aligned}
\delta {\sigma}^{ij}_s + {\tau}_{s}\frac{d \delta{\sigma}^{ij}_s}{dt} = \eta_{s} \delta {\dot{\gamma}}^{ij}, \qquad (ij=xz,\ yz),
\label{orthogonal}
\end{aligned}
\end{equation}
with
\begin{equation}
\begin{aligned}
{\tau}_{s} = \frac{\tau_{s0}}{1+\mu \dot{\gamma}_{ss}^\nu}, \qquad
{\eta}_{s} = \frac{\eta_{s0}}{1+\mu \dot{\gamma}_{ss}^\nu}.
\end{aligned}
\end{equation}
Here, $\delta \dot{\gamma}^{ij} = \gamma_0 \omega \cos \omega t$ is the shear rate of the oscillating shear strain.
Note that the equations for the $xz$ and $yz$ components are exactly identical, which arises from constitutive Eq. (\ref{conequ}) being isotropic with respect to the strain tensor $\vec{\gamma}$.
By comparing Eqs. (\ref{parallel}) and (\ref{orthogonal}), we see that the difference between the $xy$ component and the other $xz$ and $yz$ components originates from the second term of the right-hand side of Eq. (\ref{parallel}), i.e., the ``$\dot{\gamma}_{ss}^{\nu} \delta {\dot{\gamma}}^{xy}$" term.
This term arises from the coupling between the driving shear flow $\dot{\gamma}_{ss}$ and the oscillating strain $\delta \dot{\gamma}^{xy}$.
In the $xz$ and $yz$ components, such a coupling does not appear because the driving shear flow and the oscillating strain impact separate components.
However, in the $xy$ component, the driving shear flow $\dot{\gamma}_{ss}$ and the oscillating strain $\delta \dot{\gamma}^{xy}$ affect the same components, and the coupling term ``$\dot{\gamma}_{ss}^{\nu} \delta {\dot{\gamma}}^{xy}$" therefore arises, which is the origin of the difference between the $xy$ component and the other $xz$ and $yz$ components.

To conclude this section, we allude to certain studies investigating the mechanical responses of polymer solutions under a steady shear flow \cite{uneyama_2011,yamamoto_1971,wong_1989,vermant_1998,somma_2007,li_2010}, in which two responses $\delta \sigma^{xy}$ to $\delta \gamma^{xy}$ and $\delta \sigma^{yz}$ to $\delta \gamma^{yz}$ have been examined.
Note that in these studies, the responses $\delta \sigma^{xy}$ to $\delta \gamma^{xy}$ and $\delta \sigma^{yz}$ to $\delta \gamma^{yz}$ are called ``parallel superposition" and ``orthogonal superposition", respectively.
It was reported that the steady shear flow causes an effect called the convective constraint release effect \cite{somma_2007} or induces an anisotropic mobility \cite{uneyama_2011}, which both accelerate the time scale of polymer dynamics.
This result is similar to our findings that the driving shear flow makes the characteristic times faster.
However, in polymer solutions, the origin of the difference between the $xy$ and $yz$ components is still not clear.
Furthermore, to the best of our knowledge, there are no studies investigating the mechanical responses of polymer solutions for the $xz$ component.
It is expected that polymer solutions exhibit different mechanical responses in the $xz$ and $yz$ components, in contrast to supercooled liquids, which surprisingly exhibit the same responses for these components.

\section{Result II: Stress fluctuations} \label{sec:4}
In this section, the results of stress fluctuations are presented.
We show the stress correlation functions in the sheared non-equilibrium state.
In addition, we also demonstrate the violation of the FDT and present the frequency-dependent effective temperature as a metric indicating the magnitude of this violation.

\subsection{Stress correlation function} \label{sec:4-1}
\begin{figure}
\begin{center}
\includegraphics[scale=1]{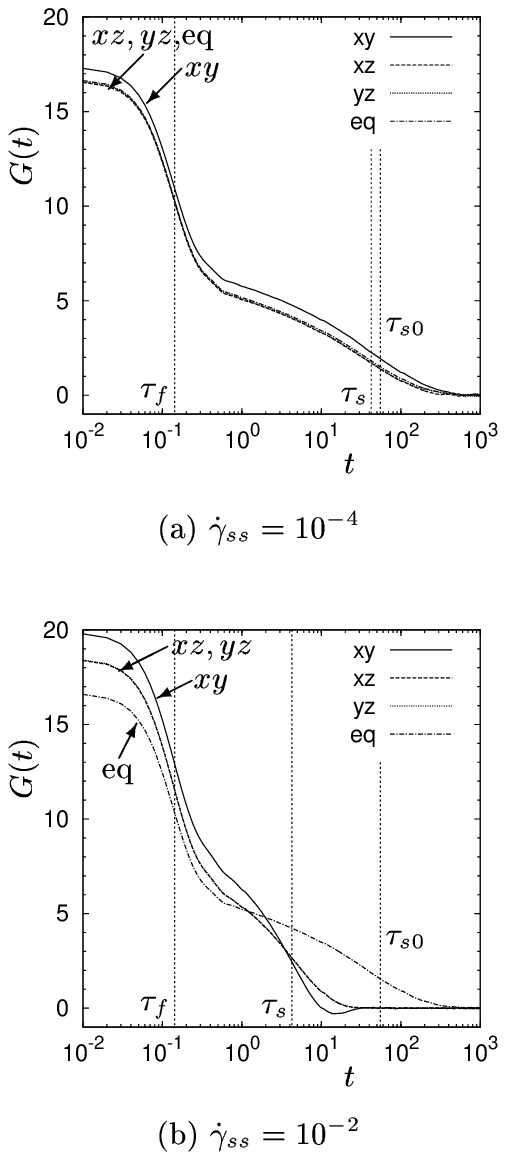}
\end{center}
\vspace*{0mm}
\caption{
The shear stress correlation function $G^{ij}_{\dot{\gamma}_{ss}}(t)$.
The temperature is $T=0.306$.
The shear rate is (a) $\dot{\gamma}_{ss}=10^{-4}$ (weakly sheared state) and (b) $\dot{\gamma}_{ss}=10^{-2}$ (strongly sheared state).
We also show $G_{\text{eq}}(t)$ in the equilibrium state.
Again, refer to the caption of Fig. \ref{result1a} for the descriptions of the labels ``$xy$", ``$xz$", ``$yz$", and ``eq" and the values $\tau_{s0}$, $\tau_s$, and $\tau_f$.
}
\label{scf}
\end{figure}

We examined the correlation function of the shear stress fluctuations, defined as
\begin{equation}
G^{ij}_{\dot{\gamma}_{ss}}(t) = \frac{V}{T}{\left< \delta \sigma^{ij}(t) \delta \sigma^{ij}(0) \right>_{\dot{\gamma}_{ss}}},
\end{equation}
where $\delta \sigma^{ij}$ represents the shear stress fluctuations in the $ij$ component of the stress tensor $\vec{\sigma}$ and $\left< \right>_{\dot{\gamma}_{ss}}$ denotes the ensemble average in the sheared non-equilibrium state.
The functions $G^{xy}_{\dot{\gamma}_{ss}}$, $G^{xz}_{\dot{\gamma}_{ss}}$, and $G^{yz}_{\dot{\gamma}_{ss}}$ are plotted in Fig. \ref{scf}, which also shows $G_{\text{eq}}$ in the equilibrium state.
The temperature is $T=0.306$.
The shear rate $\dot{\gamma}_{ss}$ is $\dot{\gamma}_{ss} = 10^{-4}$ (weakly sheared state) in Fig. \ref{scf}(a) and $\dot{\gamma}_{ss} = 10^{-2}$ (strongly sheared state) in Fig. \ref{scf}(b).
In the weakly sheared state (Fig. \ref{scf}(a)), although $G^{xy}_{\dot{\gamma}_{ss}}$ behaves slightly different from $G_{\text{eq}}$, there are only small differences between $G^{ij}_{\dot{\gamma}_{ss}}$ and $G_{\text{eq}}$; thus, the weak steady shear flow with $\dot{\gamma}_s = 10^{-4}$ produces only small effects on the shear stress fluctuations, as was observed in the mechanical responses shown in Figs. \ref{result1a} and \ref{result1b}.
However, in the strongly sheared state (Fig. \ref{scf}(b)), we can easily recognize that all $G^{ij}_{\dot{\gamma}_{ss}}$ behave much differently than $G_{\text{eq}}$.
Due to the strong shear flow, all of the $G^{ij}_{\dot{\gamma}_{ss}}$ values relax rapidly compared with $G_{\text{eq}}$.
It should be noted that even for a small time $t<\tau_f$, all $G^{ij}_{\dot{\gamma}_{ss}}$ values differ from $G_{\text{eq}}$, which means that the steady shear flow with $\dot{\gamma}_{ss}=10^{-2}$ affects the fluctuations even at $t<\tau_f \sim 0.15$, time scales that are much smaller than $\dot{\gamma}_{ss}^{-1}=10^2$.
A similar result has been observed in sheared foam \cite{ono_2002}.
This trend is different than those of the responses $G'^{ij}_{\dot{\gamma}_{ss}}$ and $G''^{ij}_{\dot{\gamma}_{ss}}$ shown in Figs. \ref{result2a} and \ref{result2b}, for which at short time scales ($\omega>\tau_f^{-1}$), all $G'^{ij}_{\dot{\gamma}_{ss}}$ and $G''^{ij}_{\dot{\gamma}_{ss}}$ values coincide with $G'_{\text{eq}}$ and $G''_{\text{eq}}$, i.e., effects due to the steady shear flow are not observed.

Fig. \ref{scf}(b) demonstrates that $G^{xz}_{\dot{\gamma}_{ss}}$ and $G^{yz}_{\dot{\gamma}_{ss}}$ are identical, whereas $G^{xy}_{\dot{\gamma}_{ss}}$ behaves quite differently and even takes on negative values.
The coincidence of $G^{xz}_{\dot{\gamma}_{ss}}$ and $G^{yz}_{\dot{\gamma}_{ss}}$ demonstrates that the shear stress in the $xz$ and $yz$ components fluctuates in the same manner despite the strong shear flow in the $xy$ component.
Such isotropic fluctuations have also been observed in the density correlation function and the mean square displacement \cite{yamamoto1_1998,miyazaki_2004,bessel_2007}.
In contrast, the different behavior of $G^{xy}_{\dot{\gamma}_{ss}}$ demonstrates an anisotropy, which has also been detected in the four-point correlation function \cite{furukawa_2009}.
Similar isotropy and anisotropy were also observed in the mechanical responses $G'^{ij}_{\dot{\gamma}_{ss}}$ and $G''^{ij}_{\dot{\gamma}_{ss}}$ depicted in Figs. \ref{result2a} and \ref{result2b}.
As discussed for Eqs. (\ref{parallel}) and (\ref{orthogonal}), the coupling between the driving shear flow $\dot{\gamma}_{ss}$ and the oscillating strain $\delta \dot{\gamma}^{xy}$ makes the response in the $xy$ component different from those in the $xz$ and $yz$ components.
We can expect that similar to the effect seen for the responses, coupling interactions between the driving shear flow and the fluctuations of the $xy$ component also arise, whereas such coupling does not appear in the $xz$ and $yz$ components.

In addition, it should be noted that the stress correlation function $G^{xy}_{\dot{\gamma}_{ss}}$ does not decay monotonically and takes on negative values at long time scales in Fig. \ref{scf}(b).
The same relaxation behavior has been also reported in sheared foam \cite{ono_2002}.
As we observed in Fig. \ref{result2a} and \ref{result2b}, the storage modulus $G'^{xy}_{\dot{\gamma}_{ss}}$ also becomes negative at long time scales (low frequencies $\omega$).
We speculate that the negative shear modulus and the negative fluctuation correlation of the xy component may be interrelated.
However, we do not yet understand the mechanism of either the negative fluctuation correlation or the negative shear modulus at this stage.

\subsection{Violation of fluctuation-dissipation theorem} \label{sec:4-2}
\begin{figure}
\begin{center}
\includegraphics[scale=1]{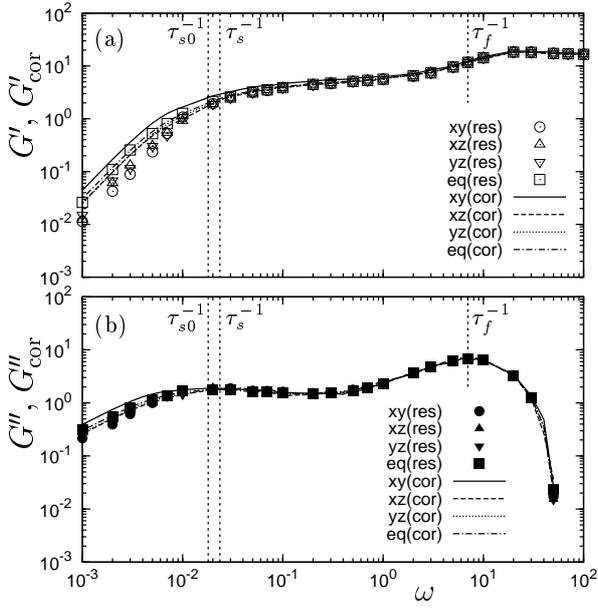}
\end{center}
\vspace*{0mm}
\caption{
The response and correlation functions (a) $G'^{ij}_{\dot{\gamma}_{ss}}$ and  $G'^{ij}_{\text{cor},\dot{\gamma}_{ss}}$ and (b) $G''^{ij}_{\dot{\gamma}_{ss}}$ and  $G''^{ij}_{\text{cor},\dot{\gamma}_{ss}}$ versus the frequency $\omega$ in the weakly sheared state $T=0.306$ and $\dot{\gamma}_{ss}=10^{-4}$.
We also show the functions in the equilibrium state.
The symbols and lines represent the response functions and the correlation functions, respectively.
The labels ``res" and ``cor" correspond to the response function and the correlation function, respectively.
Again, refer to the caption of Fig. \ref{result1a} for the descriptions of the labels ``$xy$", ``$xz$", ``$yz$", and ``eq" and the values $\tau_{s0}$, $\tau_s$, and $\tau_f$.
}
\label{result4a}
\end{figure}

\begin{figure}
\begin{center}
\includegraphics[scale=1]{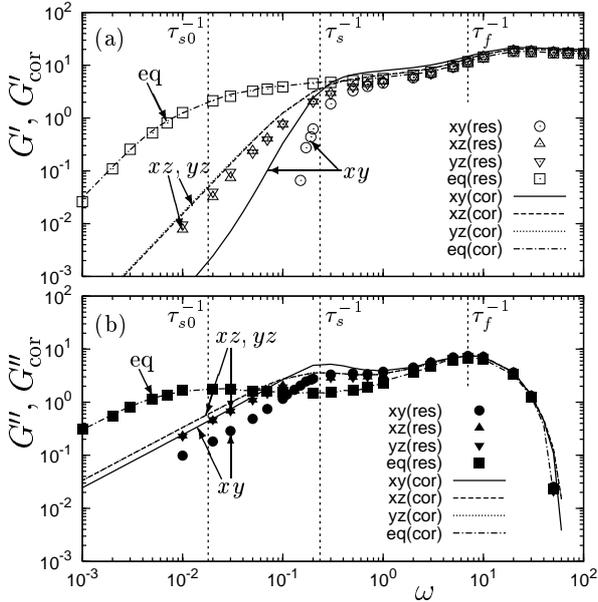}
\end{center}
\vspace*{0mm}
\caption{
The response and correlation functions (a) $G'^{ij}_{\dot{\gamma}_{ss}}$ and  $G'^{ij}_{\text{cor},\dot{\gamma}_{ss}}$ and (b) $G''^{ij}_{\dot{\gamma}_{ss}}$ and  $G''^{ij}_{\text{cor},\dot{\gamma}_{ss}}$ versus the frequency $\omega$ in the strongly sheared state $T=0.306$ and $\dot{\gamma}_{ss}=10^{-2}$.
See also the caption of Fig. \ref{result4a}.
}
\label{result4b}
\end{figure}

The relationship between the response and correlation functions is of great interest and importance \cite{harada_2005,speck_2006,chetrite_2009,baiesi_2009,baiesi_2010,uneyama_2011,berthier_2000,berthier_2002,makse_2002,ono_2002,ohern_2004,potiguar_2006,haxton_2007,kruger_2010,zhang_2011}.
If we use the shear moduli $G'^{ij}_{\dot{\gamma}_{ss}}$ and $G''^{ij}_{\dot{\gamma}_{ss}}$ as the response functions, the associated correlation functions are the Fourier transforms of the correlation function $G^{ij}_{\dot{\gamma}_{ss}}$, defined as
\begin{equation}
\begin{aligned}
G'^{ij}_{\text{cor},\dot{\gamma}_{ss}}(\omega) &= \omega \int_{0}^\infty G^{ij}_{\dot{\gamma}_{ss}}(t) \sin (\omega t) dt,\\
G''^{ij}_{\text{cor},\dot{\gamma}_{ss}}(\omega) &= \omega \int_{0}^\infty G^{ij}_{\dot{\gamma}_{ss}}(t) \cos (\omega t) dt,
\end{aligned} \label{cor}
\end{equation}
where the subscript ``cor'' denotes the ``correlation function".
We note that the response functions must be $G'^{ij}_{\dot{\gamma}_s}$ and $G''^{ij}_{\dot{\gamma}_s}$ at the small amplitude $\gamma_0=0.01$ of the oscillating strain, i.e., the linear responses.
Figures \ref{result4a} and \ref{result4b} present the correlation functions $G'^{ij}_{\text{cor},\dot{\gamma}_{ss}}$ and $G''^{ij}_{{\text{cor},\dot{\gamma}}_{ss}}$ with the response functions $G'^{ij}_{\dot{\gamma}_{ss}}$ and $G''^{ij}_{\dot{\gamma}_{ss}}$ in the weakly sheared state ($T=0.306$ and $\dot{\gamma}_{ss} = 10^{-4}$) and the strongly sheared state ($T=0.306$ and $\dot{\gamma}_{ss} = 10^{-2}$), respectively.
The functions in the equilibrium state are shown in the same figures.
We can observe that the response and correlation functions coincide with each other in the equilibrium state, which implies that the FDT holds.
In contrast, the response and correlation functions do not coincide in the non-equilibrium state, and therefore, the FDT is violated.
The large violation of FDT is observed in the strongly sheared case (Fig. \ref{result4b}).
Note that in the strongly sheared case, the FDT is violated even at high frequencies $\omega > \tau_f^{-1}$ for the storage modulus $G'^{ij}_{\dot{\gamma}_{ss}}$ and $G'^{ij}_{\text{cor},\dot{\gamma}_{ss}}$.
This fact can be more clearly observed by noting that the effective temperature $T'_{\text{eff}}$, measured from the storage modulus in Eq. (\ref{teffeq}), has slightly higher values than the temperature $T=0.306$, even at high $\omega > \tau_f^{-1}$, as shown in Fig. \ref{teffb}.
The violation of the FDT is physically interpreted as entropy production \cite{harada_2005,speck_2006}, a steady-state probability current \cite{chetrite_2009}, or other related physical quantities.
Recent works have attempted to provide a unifying framework for this phenomenon \cite{baiesi_2009,baiesi_2010}.

\subsection{Frequency-dependent effective temperature} \label{sec:4-3}
\begin{figure}
\begin{center}
\includegraphics[scale=1]{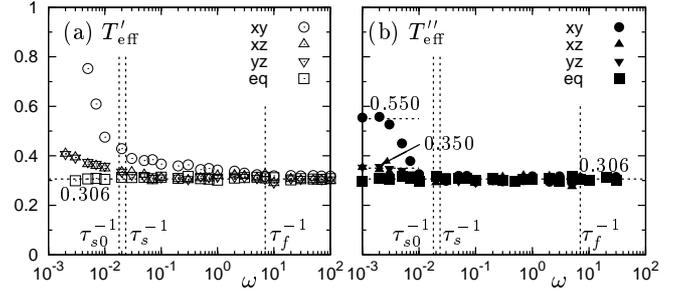}
\end{center}
\vspace*{0mm}
\caption{
The effective temperatures (a) $T'_{\text{eff}}$ and (b) $T''_{\text{eff}}$ versus the frequency $\omega$ in the weakly sheared state $T=0.306$ and $\dot{\gamma}_{ss}=10^{-4}$.
We also show the values in the equilibrium state, which coincide exactly with $T=0.306$.
Also refer to the caption of Fig. \ref{result1a} for the descriptions of the labels ``$xy$", ``$xz$", ``$yz$", and ``eq" and the values $\tau_{s0}$, $\tau_s$, and $\tau_f$.
}
\label{teffa}
\end{figure}

\begin{figure}
\begin{center}
\includegraphics[scale=1]{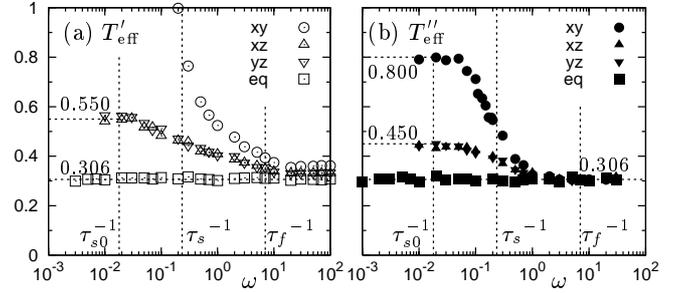}
\end{center}
\vspace*{0mm}
\caption{
The effective temperatures (a) $T'_{\text{eff}}$ and (b) $T''_{\text{eff}}$ versus the frequency $\omega$ in the strongly sheared state $T=0.306$ and $\dot{\gamma}_{ss}=10^{-2}$.
See also the caption of Fig. \ref{teffa}.
}
\label{teffb}
\end{figure}

To examine the concept of an effective temperature \cite{berthier_2000,berthier_2002,makse_2002,ono_2002,ohern_2004,potiguar_2006,haxton_2007,kruger_2010,zhang_2011}, we defined the frequency-dependent effective temperatures $T'_{\text{eff}}(\omega)$ and $T''_{\text{eff}}(\omega)$ as
\begin{equation}
\begin{aligned}
& \frac{T'_{\text{eff}}(\omega)}{T} = \frac{G'^{ij}_{\text{cor},\dot{\gamma}_s}(\omega)}{G'^{ij}_{\dot{\gamma}_s}(\omega)}, \\
& \frac{T''_{\text{eff}}(\omega)}{T} = \frac{G''^{ij}_{\text{cor},\dot{\gamma}_s}(\omega)}{G''^{ij}_{\dot{\gamma}_s}(\omega)}. \label{teffeq}
\end{aligned}
\end{equation}
Note that the effective temperatures $T'_{\text{eff}}$ and $T''_{\text{eff}}$ are defined from different observables, i.e., the storage modulus and the loss modulus, respectively.
Figures \ref{teffa} and \ref{teffb} illustrate the frequency $\omega$ dependencies of $T'_{\text{eff}}$ and $T''_{\text{eff}}$ in the weakly sheared state ($T=0.306$ and $\dot{\gamma}_{ss} = 10^{-4}$) and the strongly sheared state ($T=0.306$ and $\dot{\gamma}_{ss} = 10^{-2}$), respectively.
In the equilibrium state, $T'_{\text{eff}}$ and $T''_{\text{eff}}$ are exactly the same as the temperature $T=0.306$ for all frequencies $\omega$, as is well demonstrated in Figs. \ref{teffa} and \ref{teffb}.
Thus, the temperature $T$ exactly relates the response function to the correlation function in the equilibrium state.
However, in the non-equilibrium state, both $T'_{\text{eff}}$ and $T''_{\text{eff}}$ differ from $T=0.306$, particularly in the strongly sheared case (Fig. \ref{teffb}).
If the concept of an effective temperature is valid, then, at low frequencies $\omega$ (long time scales), $T'_{\text{eff}}$ and $T''_{\text{eff}}$ should coincide with each other and have the same value for all components $ij = xy,\ xz$, and $yz$, i.e., one scalar quantity should relate the response function and the correlation function of any observable and any component.
However, as shown in Figs. \ref{teffa} and \ref{teffb}, $T'_{\text{eff}}$ and $T''_{\text{eff}}$ differ from each other and have different values between components.
Note that in the strongly sheared state (Fig. \ref{teffb}), $T'_{\text{eff}}$ of the $xy$ component has negative values at low frequencies $\omega$ because of the negative storage modulus $G'^{ij}_{\dot{\gamma}_{ss}}$ (data are not shown in Fig. \ref{teffb}).
The differences in the effective temperatures have been also observed between different observables \cite{ono_2002,ohern_2004} and different directions \cite{zhang_2011}.
This result indicates that the use of only one effective temperature (one scalar quantity) cannot completely characterize the relationship between the response functions and the correlation functions.
For the supercooled liquid, it is notable that the effective temperatures $T'_\text{eff}$ and $T''_{\text{eff}}$ have the same values for the $xz$ and $yz$ components over all frequencies $\omega$.
This isotropic feature is considered to be a characteristic of supercooled liquids (or glassy systems).

In addition, it is interesting that the frequency $\omega$ dependencies of the effective temperatures $T'_{\text{eff}}$ and $T''_{\text{eff}}$ notably differ from each other over the entire domain of $\omega$.
The value $T''_{\text{eff}}$ coincides with $T=0.306$ at high frequencies $\omega$ \hspace{-0.1em}\raisebox{0.4ex}{$>$}\hspace{-0.75em}\raisebox{-.7ex}{$\sim$}\hspace{-0.1em} $\tau_s^{-1}$. As $\omega$ decreases, $T''_{\text{eff}}$ starts to deviate from $T=0.306$ and approaches higher values: $0.350$ for the $xz$ and $yz$ components and $0.550$ for the $xy$ component in the weakly sheared case (Fig. \ref{teffa}) and $0.450$ for the $xz$ and $yz$ components and $0.800$ for the $xy$ component in the strongly sheared case (Fig. \ref{teffb}).
This behavior of $T''_{\text{eff}}$ is consistent with previous results \cite{berthier_2000,berthier_2002,kruger_2010,zhang_2011}, i.e., the effective temperature is equivalent to the bath temperature at short times (high $\omega$) and a higher temperature at long times (low $\omega$).
However, the value $T'_{\text{eff}}$ can be slightly higher than $T$ and does not coincide with $T$ even at high $\omega > \tau_f^{-1}$, as can be clearly observed in the strongly sheared case (Fig. \ref{teffb}).
As $\omega$ decreases, $T'_{\text{eff}}$ begins to increase at a frequency that is much faster than $\tau_{s}^{-1}$ (near $\omega = \tau_f^{-1} \sim \tau_E^{-1}$ in the strongly sheared case).
In addition, $T'_{\text{eff}}$ for the $xy$ component does not approach a positive value.
Even in the weakly sheared case (Fig. \ref{teffa}), we did not obtain the asymptotic value within our simulations.
These results for $T'_{\text{eff}}$ are inconsistent with previous results \cite{berthier_2000,berthier_2002,kruger_2010,zhang_2011}.
Therefore, the effective temperature measured in previous studies \cite{berthier_2000,berthier_2002,kruger_2010,zhang_2011} may be physically similar to $T''_{\text{eff}}$ but meaningfully different from $T'_{\text{eff}}$.

\begin{figure}
\begin{center}
\includegraphics[scale=1]{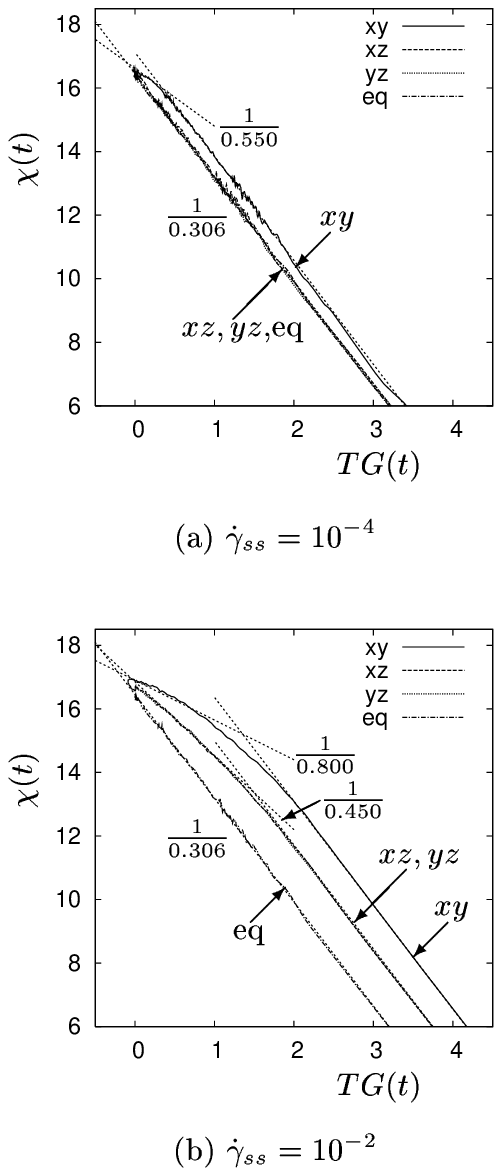}
\end{center}
\vspace*{0mm}
\caption{
The susceptibility $\chi^{ij}_{\dot{\gamma}_{ss}}(t)$ versus the shear stress correlation function $G^{ij}_{\dot{\gamma}_{ss}}(t)$.
The temperature is $T=0.306$.
The shear rate is (a) $\dot{\gamma}_{ss}=10^{-4}$ (weakly sheared state) and (b) $\dot{\gamma}_{ss}=10^{-2}$ (strongly sheared state).
We also show $\chi_{\text{eq}}(t)$ versus $G_{\text{eq}}(t)$ in the equilibrium state.
The labels ``$xy$", ``$xz$", and ``$yz$" denote $ij=xy$, $xz$, and $yz$, respectively, and the label ``eq" denotes the equilibrium state.
The slopes are indicated in the figures.
}
\label{stepfig}
\end{figure}

To demonstrate that the effective temperature measured in previous works \cite{berthier_2000,berthier_2002,kruger_2010,zhang_2011} is quantitatively identical to the present $T''_{\text{eff}}$, we investigated the mechanical response to a small constant shear strain.
At time $t=0$, the constant small shear strain $\delta \gamma^{ij} = 0.01$ was applied to the sheared supercooled liquid.
Note that the response is linear with respect to the shear strain $\delta \gamma^{ij} = 0.01$.
We calculated the response $\delta \sigma^{ij}$ using MD simulation and obtained the susceptibility $\chi^{ij}_{\dot{\gamma}_{ss}}(t)$ as
\begin{equation}
\chi^{ij}_{\dot{\gamma}_{ss}}(t) = \frac{\left< \delta \sigma^{ij}(t) - \delta \sigma^{ij}(0) \right>_{\dot{\gamma}_{ss}}}{\delta \gamma^{ij}},
\end{equation}
where $\left< \right>_{\dot{\gamma}_{ss}}$ denotes the ensemble average in the sheared non-equilibrium state.
The effective temperature $T_\text{eff}$ is defined as the inverse slope of the plot of the susceptibility-correlation function for long times \cite{berthier_2000,berthier_2002,kruger_2010,zhang_2011}, as given by Eq. (\ref{effectstep}).
\begin{equation}
\chi^{ij}_{\dot{\gamma}_{ss}}(t) = \frac{1}{T_\text{eff}} (TG^{ij}_{\dot{\gamma}_{ss}}(0) - TG^{ij}_{\dot{\gamma}_{ss}}(t)). \label{effectstep}
\end{equation}
Figure \ref{stepfig} shows the plot of the susceptibility-correlation function for the $ij=xy$, $xz$, and $yz$ components.
The shear rate $\dot{\gamma}_{ss}$ is $\dot{\gamma}_{ss} = 10^{-4}$ (weakly sheared state) in Fig. \ref{stepfig}(a) and $\dot{\gamma}_{ss} = 10^{-2}$ (strongly sheared state) in Fig. \ref{stepfig}(b).
In the same figure, the plot of the susceptibility-correlation function in the equilibrium state is also shown.
In the equilibrium state, the inverse slope is exactly $T=0.306$ in the whole region.
In contrast, in the sheared non-equilibrium state, the inverse slopes for long times take values higher than $T=0.306$, and these values coincide well with $T''_{\text{eff}}(\omega)$ at low frequencies $\omega$: $0.550$ for the $xy$ component in the weakly sheared case (Fig. \ref{stepfig}(a)) and $0.450$ for the $xz$ and $yz$ components and $0.800$ for the $xy$ component in the strongly sheared case (Fig. \ref{stepfig}(b)).
Note that in the weakly sheared case, the inverse slopes for the $xz$ and $yz$ components should be $0.350$ ($T''_{\text{eff}}(\omega)$ at low $\omega$), although we cannot distinguish such slopes numerically.
Thus, we can conclude that the effective temperature $T_\text{eff}$ used in previous works \cite{berthier_2000,berthier_2002,kruger_2010,zhang_2011} and $T''_{\text{eff}}$ are exactly same.
In fact, the effective temperatures $T_\text{eff}$ and $T''_{\text{eff}}$ are mathematically connected, as the responses to a small constant strain and an oscillating strain are related by a Fourier transformation.

\section{Conclusion} \label{sec:5}
In this study, we examined the shear stress responses and fluctuations of a supercooled liquid in sheared non-equilibrium states.
Interestingly, for the two components differing from that of the driving shear flow, the same responses and fluctuation correlations were observed despite the presence of that driving shear flow.
From these responses and fluctuations, identical magnitudes of the violations of the FDT were obtained.
These results demonstrate the isotropic aspect of this system \cite{yamamoto1_1998,miyazaki_2004,bessel_2007}.
Based on this isotropic feature, we successfully constructed the two mode-Maxwell model with isotropic non-linearity, written as Eq. (\ref{conequ}) and Eq. (\ref{param}), for the constitutive equation of supercooled liquids.
This simple constitutive equation is surprisingly accurate at describing the mechanical properties of the supercooled liquid not only in the quiescent equilibrium state but also in the sheared non-equilibrium state.

In contrast, for the same component as that of the driving shear flow, both the responses and fluctuations are quite different from those of the other two components.
This result demonstrates the anisotropic aspect of the system \cite{furukawa_2009}.
Using the constitutive equation, we demonstrated that the anisotropy in the responses derives from the coupling between the steady shear flow and the oscillating shear strain, a coupling that only arises when both factors impact the same component.
Coupling with the driving shear flow can also be expected for the stress fluctuations, which may cause anisotropic fluctuations, just as such a coupling caused anisotropic responses.
This coupling between the driving shear flow and the stress fluctuations will be a future topic.
Furthermore, for this component, we observed a negative storage modulus and a negative fluctuation correlation, which may be interrelated phenomena.
The origin and relationship between these negative values of the storage modulus and fluctuation correlation will also be addressed in future study.
Due to its anisotropic responses and fluctuations, the magnitude of this component's violation of the FDT differs from those observed for the other components \cite{zhang_2011}, which indicates that a simple scalar mapping, such as the concept of an effective temperature \cite{berthier_2000,berthier_2002,makse_2002,ono_2002,ohern_2004,potiguar_2006,haxton_2007,kruger_2010,zhang_2011}, oversimplifies the true nature of supercooled liquids under shear flow.
In fact, we quantified the magnitude of this violation using the frequency-dependent effective temperature, defined as Eq. (\ref{teffeq}), and demonstrated that the effective temperature exhibits notably different values for different components.

Finally, we stress that even in the strongly sheared situation, supercooled liquids exhibit highly isotropic natures (isotropic dynamics and structure \cite{yamamoto1_1998,miyazaki_2004,bessel_2007} as well as isotropic mechanical responses and fluctuations), which cannot be expected for typical complex fluids in which the driving shear flow induces anisotropic dynamics or a structural change \cite{rheology,phasetransition}.
The constitutive equation, based on the isotropic features of supercooled liquids, is much simpler than those proposed for polymer solutions \cite{polymetric,transport,yamamoto_1971,wong_1989,vermant_1998,somma_2007,li_2010}.
These isotropic and simple features are considered to be characteristic of the supercooled liquids (or glassy systems).
However, supercooled liquids can also exhibit complicated natures, displaying anisotropic mechanical responses, anisotropic fluctuations \cite{furukawa_2009}, and negative shear modulus and negative stress correlations.
An understanding of the mechanism underlying these complicated phenomena will lead us to a better comprehension of the violations of the FDT \cite{harada_2005,speck_2006,chetrite_2009,baiesi_2009,baiesi_2010,uneyama_2011} that they involve, as well as certain consequent modifications of the concept of an effective temperature \cite{berthier_2000,berthier_2002,makse_2002,ono_2002,ohern_2004,potiguar_2006,haxton_2007,kruger_2010,zhang_2011}.
Supercooled liquids composed of spherical or low-molecular-weight molecules may be excellent materials for investigating non-equilibrium statistical mechanics.

\begin{acknowledgement}
We wish to acknowledge Prof. K. Miyazaki, Dr. K. Kim, and Dr. T. Uneyama for their useful comments.
This work was supported by the JSPS Core-to-Core Program ``International research network for non-equilibrium dynamics of soft matter".
\end{acknowledgement}

\bibliographystyle{epj.bst}
\bibliography{paper}

\begin{thebibliography}{47}
\bibitem{colloid}
For dense colloidal dispersions composed of spherical particles, isotropic
  responses are expected similar to the present model supercooled liquid.
\end{thebibliography}

\begin{table}
\caption{Simulation cases.
We performed MD simulations, varying the temperature $T$, the shear rate $\dot{\gamma}_{ss}$ of the driving shear flow, and the amplitude $\gamma_0$ of the oscillating shear strain.
The item ``Figure" indicates the figure that shows the result of each case.}
\label{tab:1}
\begin{center}
\begin{tabular}{lllll}
\hline\noalign{\smallskip}
Case                  & $T$     & $\dot{\gamma}_{ss}$ & $\gamma_0$ & Figure              \\
\noalign{\smallskip}\hline\noalign{\smallskip}
Weakly sheared        & $0.306$ & $10^{-4}$           & $0.01$     & Fig. \ref{result1a} \\
case                  &         &                     & $0.1$      & Fig. \ref{result1b} \\
\noalign{\smallskip}\hline\noalign{\smallskip}
Strongly sheared      & $0.306$ & $10^{-2}$           & $0.01$     & Fig. \ref{result2a} \\
case                  &         &                     & $0.1$      & Fig. \ref{result2b} \\
\noalign{\smallskip}\hline\noalign{\smallskip}
Different temperature & $0.267$ & $10^{-3}$           & $0.05$     & Fig. \ref{result3a} \\
case                  &         &                     & $0.2$      & Fig. \ref{result3b} \\
\noalign{\smallskip}\hline
\end{tabular}
\end{center}
\vspace*{5cm}
\end{table}

\end{document}